\documentstyle[12pt]{article}
\newcommand{\rf}[1]{(\ref{#1})}
\newcommand{\bea}{\begin{eqnarray}}
\newcommand{\eea}{\end{eqnarray}}

\renewcommand{\a}{\alpha}
\newcommand{\n}{\nu}
\newcommand{\m}{\mu}

\newcommand{\ep}{\varepsilon}
\newcommand{\e}{\epsilon}

\renewcommand{\d}{\delta}

\newcommand{\vph}{\varphi}
\newcommand{\oh}{\frac{1}{2}}

\newcommand{\ra}{\right\rangle}
\newcommand{\la}{\left\langle}

\newcommand{\cD}{{\cal D}}

\newcommand{\prt}{\partial}
\newcommand{\sqg}{\sqrt{g}}
\newcommand{\dX}{\dot{X}}
\newcommand{\noi}{\noindent}
\newcommand{\no}{\nonumber}
\def\void{}
\def\labelmark{}

\newenvironment{formula}[1]{\def\labelname{#1}
\ifx\void\labelname\def\junk{\begin{displaymath}}
\else\def\junk{\begin{equation}\label{\labelname}}\fi\junk}%
{\ifx\void\labelname\def\junk{\end{displaymath}}
\else\def\junk{\end{equation}}\fi\junk\labelmark\def\labelname{}}

{\ifx\void\labelname\def\junk{\end{array}\end{displaymath}}
\else\def\junk{\end{array}\right.\end{equation}}
\fi\junk\labelmark\def\labelname{}\def\junk{}
}

\newcommand{\beq}{\begin{formula}}
\newcommand{\eeq}{\end{formula}}
\newcommand{\beqv}{\begin{formula}{}}

\begin{document}
\topmargin 0pt
\oddsidemargin 5mm
\headheight 0pt
\headsep 0pt
\topskip 9mm

\hfill    NBI-HE-96-21

\hfill May  1996

\begin{center}
\vspace{24pt}
{\large \bf The measure in three dimensional \\
Nambu-Goto string theory}

\vspace{24pt}

{\sl J. Ambj\o rn } and {\sl A. Sedrakyan}\footnote{Permanent add
ress:
Yerevan Physics Institute, Br.Alikhanian st.2, Yerevan 36, Armenia}

\vspace{6pt}

 The Niels Bohr Institute\\
Blegdamsvej 17, DK-2100 Copenhagen \O , Denmark\\

\end{center}

\vspace{24pt}

\vfill

\begin{center}
{\bf Abstract}
\end{center}

\vspace{12pt}

\noindent
We show that the measure of the three dimensional Nambu-Goto string theory
has a simple decomposition as a measure on two parameter group of induced
area-preserving  transformations of the immersed surface and a trivial
measure for the area of the surface.

\vfill

\newpage

\noi
{\bf 1.}\\
\noi
String theory was invented to describe hadronic physics in four dimensions \cite{1}. The action is simple and beautiful: the area of the world sheet of the
string \cite{2,3}. However, canonical quantization of this so called Nambu-Goto
string is only consistent in 26 dimensions. The string theory of
Polyakov arose as an attempt to perform a consistent quantization
in lower dimensions than $d=26$ \cite{4}.
His approach can be viewed as two-dimensional
gravity coupled to $d$ scalar fields which are  identified with the
coordinates of the string in target space. The goal of a consistent
quantization in lower dimensional space was achieved, but unfortunately
only in dimensions $d \leq 1$.  A consistent quantization
of bosonic string theories in the physical
interesting dimensions $d=3$ and $d=4$  still remains to be constructed.

It is unknown if the quantum theories of the Nambu-Goto string
and the Polyakov string agree in non-critical dimensions, partly
because the path integral approach is not readily available for
the Nambu-Goto string. In this article we will investigate the
path integral which enters in the quantization of Nambu-Goto
string in $d=3$. The Nambu-Goto string invites us to work directly
with the immersed surfaces $X(\xi)$, rather than  the matter field
$X(\xi)$ taken together with an independent metric $g_{\alpha\beta}$,
as in the Polyakovs \cite{4} approach. Our aim will be to
formulate the path integral as an integral over the
surfaces
\beq{*1}
X(\xi) \equiv \Bigl( X_1(\xi_1,\xi_2),X_2(\xi_1,\xi_2),X_3(\xi_1,\xi_2)\Bigr).
\eeq
We will show that the generators of infinitesimal deformations of
$X$ which
leave the Nambu-Goto Lagrangian  invariant form a two-parameter
local Lie algebra, which can be considered as an algebra of vector
 fields on the
manifold of maps \cite{5,6} of the two-dimensional world sheet into
the three-dimensional target space $R^3$. This "induced"
area preserving algebra
contains the ordinary area preserving algebra of two-dimensional
manifolds as a subalgebra.  Further, the measure $\cD X $ of
integration over surfaces is equal to the measure for the
area of the surface, $\cD \ln \sqrt{g}$, times  the measure $\cD \e_\a$,
where $g_{ab}(\xi)$ is the metric induced on $X$ in $R^3$ and
$\e_a$ are the parameters of the above mentioned new symmetry
group. No Jacobian appears and we will be left with the following
expression for the partition function:
\beq{*2}
Z(\a_0) = \int \cD X(\xi) \; e^{-S_{NG} (X)} = \int \cD \ln \sqrt{g}(\xi)
\; e^{-\frac{1}{\a_0}\int d^2\xi \sqg} \cD \ep_\a(\xi) ,
\eeq
where $1/\a_0$ is the string tension and
\beq{*3}
S_{NG} (X) = \frac{1}{\a_0} \int d^2\xi \; \sqrt{ (\prt_a X^\m(\xi))^2
(\prt_b X^\n(\xi))^2 - (\oh \ep^{ab}\prt_a X^\m(\xi) \prt_b X^\m(\xi))^2}.
\eeq

\vspace{24pt}

\noi
{\bf  2.}\\
\noi
Let us first consider the analogous problem for the particle
in two dimensions. The action of a particle is the length
of the world line:
\beq{*4}
S(X) = m \int_0^1 d\xi \;e(\xi),~~~~~~
e(\xi ) \equiv \sqrt{{\dot{X}_\m(\xi)}^2}
\eeq
where $\dot{X}(\xi)$ denotes differentiation with respect to the
parameter $\xi$, which we assume takes values in $[0,1]$.
Let us consider the transformations of paths
$$
\xi \mapsto X(\xi)\equiv \Bigl(X_1(\xi),X_2(\xi)\Bigr)
$$
which preserve the induced metric $e(\xi)$.
The general form  of an infinitesimal transformati
on is
\beq{*5}
\d X(\xi)  = \e (\xi) \dX (\xi) + B(\xi) N(\xi),
\eeq
where $N(\xi)$ is the normal to the curve $X$ and $\e$ and $B$ are
infinitesimal. A length-preserving transformation $\d_g X$ satisfies
\beq{*6}
\d e(\xi) =0,
\eeq
and one easily checks that it is equivalent to
\beq{*7}
B_e(\xi) = \frac{1}{k(\xi)} D_\xi \e,
\eeq
where $D_\xi \equiv e^{-1}\dot{e}$ is the covariant
derivative with respect to the metric and
$k(\xi)$ is the curvature of the trajectory $X(\xi)$, i.e.
$$
k(\xi) 
= \frac{\ep^{\m\n}\dX_\m \ddot{X}_\n}{e^3}.
$$

In addition one can show that the transformations \rf{*5}
with the imposed constraint \rf{*7},
\beq{*8}
\d_e X = \e \,\dX + \frac{D_\xi \e}{k}  \; N,
\eeq
form an algebra with the commutation relations
\beq{*9}
\d_2 (\d_1 X) - \d_1 (\d_2 X) = \d_{(1,2)} X,
\eeq
where
\beq{*10}
\e_{(1,2)} = \e_1 \dot{\e}_2 - \e_2 \dot{\e}_1 -
\frac{1}{(k e)^2} \Bigl( \dot{\e}_1 \ddot{\e}_2-
\dot{\e}_2 \ddot{\e}_1\Bigr).
\eeq

The measure  $\cD X(\xi)$ can be 
transformed to a
measure   $\cD \ln e(\xi) \cD \e(\xi) $. The metric on the tangent
space of maps $X : [0,1] \mapsto R^2$ is defined by
\beq{*11}
\la \d X ,\d X\ra = \int_0^1  e(\xi) d \xi\; \d X(\xi) \cdot \d X(\xi).
\eeq
It is possible to decompose the deformations $\d X$ into
the deformations $\d_e X$ given by \rf{*8}, which are
area preserving and the deformations $\d_\bot X$ orthogonal to these
in the metric \rf{*11}. From
\beq{*12}
\la\d_e X, \d_\bot X\ra =0
\eeq
one obtains
\beq{*13}
\d_\bot X = - \frac{\dot{B}_\bot}{e^2} \, \dX + B_\bot \, N,
\eeq
where $B_\bot(\xi)$ is a free parameter.
While the deformation $\d_e X$ is characterized by
the property $e(X(\xi)) =e (X(\xi)+\d_e X(\xi))$ the deformations $\d_\bot X$
will result in  a change in $e(\xi)$. One finds
\beq{*14}
\d \ln e = -\Bigl[ k^2 - \Bigl(\frac{1}{e}\frac{d}{d\xi}\Bigr)^2\Bigr]
\frac{B_\bot}{k} .
\eeq
This implies that the transformation from  $B_\bot/k$ to $\d e$
has a Jacobian
\beq{*15}
\det \Bigl[ k^2 - \Bigl( \frac{1}{e}\frac{d}{ d\xi} \Bigr)^2 \Bigr]^{-1}.
\eeq
In addition we have
\bea
\la  \d X,\d X \ra &=& \int_0^1 e\,d\xi \,
\left(\Bigl(\e \dX + \frac{1}{k} D_\xi \e \, N\Bigr)^2+
\Bigl( -\frac{\dot{B}_\bot}{e^2} \, \dX + B_\bot \, N\Bigr)^2\right)
\label{*16}\\
& = & \int_0^1 e\,d\xi \,\left( e \e \Bigl[ 1 - \frac{1}{e}\frac{d}{ d\xi}
\frac{1}{k^2 e}\frac{d}{d\xi} \Bigr] e \e +
\frac{B_\bot}{k} \Bigl[ \Bigl(\frac{1}{e}\frac{d}{d\xi}\Bigr)^2 -k^2
\Bigr] \frac{B_\bot}{k} \right).               \no
\eea
From \rf{*14} and \rf{*16} it follows that
\beq{*17}
\cD X  = \cD \ln e \, \cD ( e \e/k ).
\eeq
This formula shows that there is no Jabobian related to the
transformation
\beq{*18}
\d X \to \Bigl( \d e /e,  e \e/k\Bigr).
\eeq

The geometric interpretation of the variable $ k e \e $ is seen if we
use the proper time parametrization of the path $X(\xi)$. If $t$
denotes the proper time
$$
 e \e = dt = \frac{dt}{d \phi} d \phi = \frac{1}{k} d \phi.
$$
and the integration over $k e\e $ c
an be viewed as the
integration over  successive angles $\phi(t)$ of the
tangents of paths. Of course one could have
introduced this decomposion of $\cD X$ from the beginning.
However, the derivation of \rf{*17} presented above has
the virtue that it allows generalization to surfaces immersed in $R^3$.

\vspace{24pt}

\noi
{\bf  3.}\\
\noi
Let us generalize the results of the last section to surfaces
immersed in $R^3$.  Let $M_2$ be a two-dimensional manifold
and $X : M_2 \mapsto R^3$ the map 
to $R^3$.
We use the  notation
\beq{*22}
g_{ab} = \prt_a X^\mu \prt_b X^\mu ,~~~~~\det g_{ab} = g,
\eeq
for the induced metric  and its determinant, respectively.

A general deformation of the surface $X(\xi)$ can be written as
\beq{*21}
\d X = \e^a \prt_a X + B N
\eeq
where $\prt_a X$ denote  the two tangents and $N$ the normal to $X$.
The deformation $\d_g X$ is area preserving if $\d g=0$, which implies
\beq{*23}
B_g = \frac{1}{h} D_a \e^a,
\eeq
where $D_a$ is the covariant derivative with respect to the
induced metric $g_{ab}$, $\e^a_g$ are free parameters  and
\beq{*24}
h_{ab} = N \cdot D_a \prt_b X,~~~~~h = h_{ab}g^{ab}
\eeq
are the second fundamental form and the mean curvature of $X$, respectively.

As in the particle case one can check that deformations \rf{*21}
with the constraint \rf{*23}, i.e. deformations
\beq{*25a}
\d_g X =\e^a \prt_a X+\frac{1}{h} D_a \e^a \, N
\eeq
form a Lie algebra:
\beq{*25}
\d_2(\d_1 X) - \d_1(\d_2 X) = \d_{(1,2)} X,
\eeq
where
\beq{*26}
\
e^a_{(1,2)} = \e^b_1 D_b \e_2^a - \e_2^b D_b \e_1^a +
\Bigl(\frac{D_c \e_2^c}{h}\Bigr) \prt^a \,\Bigl(\frac{D_b \e_1^b}{h}\Bigr)  -
\Bigl(\frac{D_c \e_1^c}{h}\Bigr) \prt^a \,\Bigl(\frac{D_b \e_2^b}{h}\Bigr) .
\eeq
The first two terms correspond to the usual terms present in
the Lie algebra for vector fields on the tangent space of $X$
and if $D_a \e^a =0$ we have according to \rf{*25a}
the standard area preserving diffeomorphism algebra.

The measure $\cD X$ is defined on the tangent space
of the 
maps $X: M^2 \mapsto R^3$ by
\beq{*23a}
\la \d X, \d X\ra = \int_{M_2} \sqg d^2\xi \;\d X(\xi) \cdot \d X(\xi),
\eeq
as in the case of the particle. Again we can split the deformations
\rf{*21} in two classes: $\d X_g$ which leaves $\sqrt{g(\xi)}$ invariant and
where $B_g$ satisfy \rf{*23}, and $\d X_\bot$, orthogonal
to $\d X_g$ with respect to the scalar product \rf{*23a}. In analogy
with the particle case \rf{*13} one finds from
\beq{*24a}
\la \d_g X,\d_\bot X\ra =0
\eeq
that
\beq{*26a}
\e^
a_\bot = \prt^a (B_\bot/h),
\eeq
for  the deformation $\d_\bot X = \e^a_\bot \prt_a X+ B_\bot N$.
With this decomposition the norm of an arbitrary tangent
vector on the space of maps $X : M_2 \mapsto R^3$ is
\bea
\la \d X,\d X \ra & =&  \int_{M_2} \sqg d^2\xi \Bigl( \d_g X
\cdot \d_g X +
\d_\bot X \cdot\d_\bot X\Bigr) \label{*27}\\
&=& \int_{M_2} \sqg d^2\xi \;
\e^a \Bigl[ \d_{ab} - D_a \Bigl( \frac{1}{h^2} D_b\Bigl) \Bigr] \e^b+
B_\bot \Bigl[ 1- \frac{1}{h} D_a \prt^a\frac{1}{h} \Bigr] B_\bot. \no
\eea
An arbitrary two-dimensional vector $\e^a$ can be represented as
\beq{*28}
\e^a = \prt^a \vph + \frac{1}{\sqg}\ep^{ab}\prt_b \phi.
\eeq
and we have
\beq{*29}
\int_{M_2} \sqg d^2\xi  \; \e^a \e_a = -
\int_{M_2} \sqg d^2\xi
\; \Bigl( \vph \nabla^2 \vph + \phi \nabla^2 \phi\Bigr)
\eeq
from which we deduce
\beq{*30}
\cD \e^a = \det (-\nabla^2) \; \cD\vph \,\cD \phi.
\eeq
From \rf{*27} we now obtain
\beq{*31}
\la \d X,\d X\ra = \int_{M_2} \sqrt{g} d^2\xi
\Bigl( \vph (-\nabla^2 + \nabla
^2 \frac{1}{h^2} \nabla^2) \phi +
\phi( -\nabla^2) \phi +
B_\bot(1- \frac{1}{h} \nabla^2 \frac{1}{h}) B_\bot \Bigr).
\eeq
Hence,
\beq{*32}
\cD X = \det\sqrt{-\nabla^2} \det\sqrt{-\nabla^2
(h^2- \nabla^2)}\det h^{-1} \det\sqrt{-\nabla^2+h^2}\; \cD \vph\,\cD \phi\,
\cD \Bigl(\frac{B_\bot}{h}\Bigr),
\eeq
and by the use of \rf{*30} we can write $\cD X$ as
\beq{*33}
\cD X = \det (h^2 -\nabla^2) \; \cD \Bigl(\frac{\e^a}{h}\Bigr) \,
\cD \Bigl(\frac{B_\bot}{h}\Bigr).
\eeq
As in the particle case we  can use $\d g$ instead of $B_\bot$.
One has
\beq{*34}
\d \ln \sqg = (D_aD^a - h^2) \Bigl( \frac{B_\bot}{ h}\Bigl),
\eeq
i.e.
\beq{*35}
\cD \ln \sqg = \det (h^2- D_aD^a) \; \cD \Bigl(\frac{B_\bot}{h}\Bigr).
\eeq
Using \rf{*34} and \rf{*35} we can finally write $\cD X$ as
\beq{*36}
\cD X = \cD \ln \sqg \; \cD \e^a,
\eeq
and the partition function for the Nambu-Goto string theory reads
\beq{*37}
Z(\a_0) = \int \cD \ln \sqg \;e^{-\frac{1}{\a_0} \int_{M_2} \sqrt{g}d^2\xi }
\int \cD \e^a
\eeq

\vspace{24pt}

\noi
{\bf  4.}\\
\noi
The immediate question which arises is in what kind of mathematical framework
one should view the Lie algebra defined by eqs. (23)-(25).
We here draw the attention to the work in \cite{5} (see \cite{7}) for 
a review) which is a generalization of the classical work of Lie 
on transformation groups.


The main idea is to introduce the so called
the prolongation of the ordinary Lie-B\"{a}cklund operator
\begin{equation}
L_0=\epsilon^a\frac{\partial}{\partial\xi^a}+B N^\mu
\frac{\partial}{\partial x^\mu}
\end{equation}
by derivatives of the map $X^\mu(\xi)$ considering the derivatives  as
independent variables
\begin{equation}
L=\epsilon^a\frac{\partial}{
\partial\xi^a}+
B N^\mu
\frac{\partial}{\partial x^\mu}+
\Xi^\mu_a
\frac{\partial}{\partial (\partial_a x^\mu)}\;,
\end{equation}
where
\begin{equation}
\Xi^\mu_a= \partial_a(B N^\mu)
-\partial_a(\epsilon^b\partial_b X^\mu)+
B\partial_a\partial_b X^\nu\frac{\partial N^\mu}
{\partial(\partial_b X^\nu)}
\end{equation}
The first two terms in eq.(41) appeared due to definition of
$\delta X^\mu(\xi)$ in
eq.(23) whereas the  next term constitutes the prolongation. 
The invariance condition
\begin{equation}
L\sqrt{g}=0
\end{equation}
is called the determining equation for the Lie-B\"{a}cklund operator and
it fixes the constraint  on B to be of the form (21).
It is easily checked that Lie bracket algebra defined by prolonged operator
(41) forms the algebra we found in eq.(25).

\vspace{24pt}

\noi
{\bf 5.}\\
\noi
We have shown that the measure $\cD X$ can be written as
$\cD \ln \sqg \,\cD \e^a$, without any determinant.  The parameters
$\e^a$, as introduced here, were related to what we call
``induced'' area preserving deformations of the surface $X(\xi)$.
These deformations form a continuous group and we showed by
explicit calculation that the generators of deformations
form a Lie algebra. The group contains as a subgroup the
usual group of area preserving diffeomorphisms, but in addition
it contains the deformations of $X$ in $R^3$ which preserve the
area density $\sqg$ and which describe the fluctuations
of the surface viewed
 as an incompressible membrane in $R^3$.
The additional part of the measure, including the area action,
is a simple product measure.

The Nambu-Goto action, as well as all our transformations (23) are
reparametrization invariant and therefore the physical space of states
should be defined in the factor space of maps $\{\epsilon^a, \sqrt{g}\}$
over diffeomorphisms. The physics will be encoded in this factor space,
and it is natural to expect that the fluctuations of the incompressible
membrane in $R^3$ will play an important role in describing this physics.
In the case of the particle this is manifest in the sense that one can 
first perform the integration over all positions of the world--line 
with two endpoints fixed and a given length and only afterwards the 
integration over world--lines of different length. In this way one 
recovers the free massive propagator of a particle \cite{adj}. Since the 
factorization of the path integral of the three-dimensional string is
very similar to the factorization of the path integral for the two-dimensional
particle, it is natural to conjecture that a similar decomposition 
will take place for the string and that it should be possible 
to perform first the integration over induced area preserving deformations
when we calculate for instance two-point correlators of the bosonic string.

\vspace{24pt}

\noi
{\bf Acknowledgement}\\
\noi
The authors thanks B.Durhuus, J.Petersen for interesting discussions.
This work supported in part by INTAS Grant 94-840.
JA acknowledge the support of the Professor Visitante Iberdrola Grant and
the hospitality at the University of Barcelona, where part of
this work were done.

\vspace{24pt}


\end{document}